\newcommand{\be}{\begin{equation}}\newcommand{\ee}{\end{equation}}
\newcommand{\bea}{\begin{eqnarray}}\newcommand{\eea}{\end{eqnarray}}
\newcommand{\p}[1]{(\ref{#1})}
\documentstyle[12pt]{article}

\begin{document}
\renewcommand{\thefootnote}{\fnsymbol{footnote}}
\thispagestyle{empty}
\begin{center}
{\hfill  hep-th/9503118   }\vspace{2cm} \\
MORE ON THE LINEARIZATION OF $W$-ALGEBRAS \vspace{1cm} \\
S. Krivonos\footnote{E-mail: krivonos@thsun1.jinr.dubna.su} and
A. Sorin\footnote{E-mail: sorin@thsun1.jinr.dubna.su} \vspace{1cm}\\
Bogoliubov Laboratory of Theoretical Physics, JINR,\\
141980, Dubna, Moscow Region, Russia \vspace{3cm} \\
{\bf Abstract}
\end{center}
We show that a wide class of $W$-(super)algebras, including
$W_N^{(N-1)}$, $U(N)$-superconformal as well as $W_N$
nonlinear algebras, can be linearized by embedding them as
subalgebras into some {\em linear} (super)conformal algebras with finite sets
of currents. The general construction is illustrated by the example of $W_4$
algebra.\vspace{1cm}\\
\begin{center}
{\it Submitted to Phys. Lett. B}
\end{center}
\vfill
\setcounter{page}0
\renewcommand{\thefootnote}{\arabic{footnote}}
\setcounter{footnote}0
\newpage

\section{Introduction}

Since the pioneer paper of Zamolodchikov \cite{W3}, a lot of extended
nonlinear conformal algebras (the $W$-type algebras) have been constructed
and studied (see, e.g., \cite{BS} and references therein). The growing
interest to this subject is motivated by many interesting applications of
nonlinear algebras to the string theory, integrable systems, etc.
However, the intrinsic nonlinearity of W-algebras makes it rather difficult
to apply to them the standard arsenal of techniques and means used in the
case of linear algebras (while constructing their field representations,
etc.). A way to circumvent this difficulty has been proposed by us
in \cite{KS}. We found that in many cases a given nonlinear $W$ algebra
can be embedded into some linear conformal algebra which is generated by a
finite number of currents and contains the considered $W$-algebra as
subalgebra in some nonlinear basis.
Up to now the explicit construction has been carried out for
some simplest examples of nonlinear (super)algebras ( $W_3$ and $W_3^{(2)}$
\cite{KS}, $WB_2$ and $W_{2,4}$ \cite{BKS} ).
Besides being a useful tool to construct new field realizations
of nonlinear algebras [3-4], these linear algebras provide
a suitable framework for considering the embeddings of the Virasoro string in
the $W$-type ones \cite{BO}.

In the present letter\footnote{The preliminary version of this Letter
has been present as talk at the International Workshop "Finite Dimensional
Integrable Systems", July 18-21, JINR, Dubna, 1994.}
 we show that the linearization is a general property
inherent to many nonlinear $W$-type algebras. We demonstrate that a
wide class of $W$-(super)algebras, including
$U(N)$-superconformal \cite{KB}, $W_N^{(N-1)}$ [7-9],
as well as $W_N$ \cite{FL} algebras, admit a linearization.
The explicit formulas related linear and nonlinear algebras for all these
cases are given. The example of $W_4$ algebra is elaborated in detail.

\setcounter{equation}0
\section{Linearizing $U(N)$ (quasi)superconformal algebras.}

In this Section we construct linear conformal algebras which
contain the algebra $W_{N+2}^{(N+1)}$ or $U(N)$ superconformal algebras
as subalgebras in some nonlinear basis.
By this we mean, that the currents of nonlinear
algebras can be related by an {\it invertible} transformation to those of
linear algebras.
In what follows these linear algebras will be called the linearizing algebras
for nonlinear ones.

Let us start by reminding the operator product expansions (OPE's) for the
$W_{N+2}^{(N+1)}$ algebras and $U(N)$ superconformal algebras (SCA).
The OPE's for these algebras can be written in a general uniform way
keeping in mind that the $W_N^{(N-1)}$ algebra is none other than $U(N-2)$
quasi-superconformal algebra (QSCA) [7-9] \footnote{Strictly speaking, the
$W_N^{(N-1)}$ algebra coincides with $GL(N-2)$ QSCA. In what follows,
we will not specify the real forms of algebras and use the common
term $U(N)$ QSCA.}.
Both $U(N)$ SCA and $U(N)$ QSCA  have
the same number of generating currents: the stress tensor $T(z)$,
the $U(1)$ current $U(x)$, the $SU(N)$  Kac-Moody currents $J_{a}^{b}(x)$
$(1\leq a,b \leq N , \mbox{Tr}(J)=0)$ and two sets of
currents in the fundamental $G_a(x)$ and conjugated ${\bar G}^b(x)$
representations of $SU(N)$. The currents $G_a(x),{\bar G}^b(x)$ are bosonic
for $U(N)$ QSCA and fermionic for $U(N)$ SCA. To distinguish between
these two cases we, following refs. \cite{Rom}, introduce the parameter
$\epsilon$ equal to $1 (-1)$ for the QSCAs (SCAs)  and write
the OPE's for these algebras in the following universal form:
\bea
T(z_1)T(z_2) & = & \frac{c/2}{z_{12}^4}+\frac{2T}{z_{12}^2}+
                   \frac{T'}{z_{12}} \quad , \quad
U(z_1)U(z_2)  =  \frac{c_1}{z_{12}^2} \; , \nonumber \\
T(z_1)J_{a}^{b}(z_2) & = & \frac{J_{a}^{b}}{z_{12}^2}+
                   \frac{{J_{a}^{b}}'}{z_{12}} \quad , \quad
T(z_1)U(z_2)  =  \frac{U}{z_{12}^2}+\frac{U'}{z_{12}} \;, \nonumber \\
T(z_1)G_a(z_2)& = & \frac{3/2 G_a}{z_{12}^2}+\frac{G_a}{z_{12}}\quad , \quad
T(z_1){\bar G}^a(z_2) =  \frac{3/2 {\bar G}^a}{z_{12}^2}+
                 \frac{{\bar G}^a}{z_{12}}\; , \nonumber \\
J_a^b(z_1)J_c^d(z_2) & = & (K-\epsilon -N)\frac{\delta_a^d\delta_c^b-
 \frac{1}{N}\delta_a^b\delta_c^d}{z_{12}^2}+
 \frac{\delta_c^b J_a^d-\delta_a^d J_c^b}{z_{12}} \; , \nonumber \\
U(z_1)G_a(z_2) & = & \frac{G_a}{z_{12}} \quad , \quad
U(z_1){\bar G}^a(z_2)  =  -\frac{{\bar G}^a}{z_{12}} \; , \nonumber \\
J_a^b(z_1)G_c(z_2) & = &
 \frac{\delta_c^b G_a -\frac{1}{N}\delta_a^b G_c}{z_{12}} \quad , \quad
J_a^b(z_1){\bar G}^c(z_2)  =  \frac{-\delta_a^c {\bar G}^b +
 \frac{1}{N}\delta_a^b {\bar G}^c}{z_{12}} \; \nonumber \\
G_a(z_1) {\bar G}^b(z_2) & = & \frac{2\delta_a^b c_2}{z_{12}^3}+
 \frac{2x_2\delta_a^b U + 2x_3 J_a^b}{z_{12}^2}+
 \frac{x_2\delta_a^b U' + x_3 {J_a^b}'+2x_5 (J_a^dJ_d^b)}{z_{12}}+
          \nonumber \\
 & & \frac{2x_4 (UJ_a^b)+\delta_a^b \left( x_1 (UU)- 2\epsilon T +
     2x_6 (J_d^eJ_e^d)\right) }{z_{12}} \; , \label{ope}
\eea
where the  central charges $c$ and parameters $x$ are defined by
\bea
c & = & \frac{-6\epsilon K^2 +(N^2+11\epsilon N +13)K-(\epsilon+N)
     (N^2+5\epsilon N+6)}{K} \; , \nonumber \\
c_1 & = & \frac{N(2K-N-2\epsilon)}{2+\epsilon N} \quad , \quad
  c_2=\frac{(K-N-\epsilon )(2K-N-2\epsilon )}{K} \; ,  \nonumber \\
x_1 & = & \frac{(\epsilon +N)(2\epsilon+N)}{N^2K} \quad , \quad
 x_2 = \frac{(2\epsilon +N)(K-\epsilon-N)}{\epsilon NK} \quad , \quad
 x_3 = \frac{2K-N-2\epsilon }{K} \; , \nonumber \\
x_4 & = & \frac{2+\epsilon N}{NK} \quad , \quad
 x_5 = \frac{1}{K} \quad , \quad x_6 = \frac{1}{2\epsilon K} \; .
            \label{opecoeff}
\eea
The currents in the r.h.s. of OPE's \p{ope} are evaluated at the point $z_2$,
$z_{12}=z_1-z_2$ and the normal ordering in the nonlinear terms is
understood.

The main question we need to answer in order to linearize the
algebras \p{ope} is as to which minimal set of additional currents must be
added to \p{ope} to get  extended linear conformal algebras
containing \p{ope} as subalgebras. The idea of our construction comes
>from the observation that the classical $(K \rightarrow \infty )$
$U(N)$ (Q)SCA \p{ope} can be realized as left shifts
in the following coset space
\be
g = e^{\int \! dz {\bar Q}^a (z) G_a (z) } \;,\label{coset}
\ee
which is parametrized by  $N$ parameters-currents ${\bar Q}^a(z)$ with
unusual conformal weights $-1/2$. In this case, all the currents of
$U(N)$ (Q)SCA \p{ope} can be constructed from  ${\bar Q}^a(z)$,
their conjugated  momenta $G_a(z) = \delta /\delta {\bar Q}^a$
and the currents of the maximal {\it linear} subalgebra ${\cal H}_N$
\be
{\cal H}_N = \left\{ T , U, J_a^b , {\bar G}^a \right\} \; . \label{glin}
\ee
Though the situation in quantum case is more difficult, it seems still
reasonable to try to extend the $U(N)$ (Q)SCA  \p{ope} by $N$
additional currents  ${\bar Q}^a(z)$ with conformal weights
$-1/2$.\footnote{Let us remind that the current with just this conformal
weight appears in the linearization of $W_3^{(2)}$ algebra \cite{KS}.}

Fortunately, this extension is sufficient to construct the
linearizing algebras for the $U(N)$ (Q)SCAs.
Without going into details, let us write down  the set of OPE's for
these linear algebras, which we will denote as $(Q)SCA_N^{lin}$
\bea
T(z_1)T(z_2) & = & \frac{c/2}{z_{12}^4}+\frac{2T}{z_{12}^2}+
                   \frac{T'}{z_{12}} \quad , \quad
U(z_1)U(z_2)  =  \frac{c_1}{z_{12}^2} \; , \nonumber \\
T(z_1)J_{a}^{b}(z_2) & = & \frac{J_{a}^{b}}{z_{12}^2}+
                   \frac{{J_{a}^{b}}'}{z_{12}} \quad , \quad
T(z_1)U(z_2)  =  \frac{U}{z_{12}^2}+\frac{U'}{z_{12}} \;, \nonumber \\
T(z_1)G_a(z_2)& = & \frac{3/2 G_a}{z_{12}^2}+\frac{G_a}{z_{12}}\quad , \quad
T(z_1)\widetilde{\overline G}{}^a(z_2) =
     \frac{3/2 \widetilde{\overline G}{}^a}{z_{12}^2}+
                 \frac{\widetilde{\overline G}{}^a}{z_{12}}\; , \nonumber \\
T(z_1){\bar Q}^a(z_2) & = & \frac{-1/2 {\bar Q}^a}{z_{12}^2}+
                 \frac{{\bar Q}^a}{z_{12}}\; , \nonumber \\
J_a^b(z_1)J_c^d(z_2) & = & (K-\epsilon -N)\frac{\delta_a^d\delta_c^b-
 \frac{1}{N}\delta_a^b\delta_c^d}{z_{12}^2}+
 \frac{\delta_c^b J_a^d-\delta_a^d J_c^b}{z_{12}} \; , \nonumber \\
U(z_1)G_a(z_2) & = & \frac{G_a}{z_{12}} \quad , \quad
U(z_1)\widetilde{\overline G}{}^a(z_2)  =
     -\frac{\widetilde{\overline G}{}^a}{z_{12}}  \quad , \quad
U(z_1){\bar Q}^a(z_2)  =  -\frac{{\bar Q}^a}{z_{12}} \; , \nonumber \\
J_a^b(z_1)G_c(z_2) & = &
 \frac{\delta_c^b G_a -\frac{1}{N}\delta_a^b G_c}{z_{12}} \quad , \quad
J_a^b(z_1)\widetilde{\overline G}{}^c(z_2)  =
       \frac{-\delta_a^c \widetilde{\overline G}{}^b +
 \frac{1}{N}\delta_a^b \widetilde{\overline G}{}^c}{z_{12}} \; , \nonumber \\
J_a^b(z_1){\bar Q}^c(z_2) & = & \frac{-\delta_a^c {\bar Q}^b +
 \frac{1}{N}\delta_a^b {\bar Q}^c}{z_{12}} \; , \nonumber \\
G_a(z_1) {\bar Q}^b(z_2) & = & \frac{\delta_a^b}{z_{12}}
 \quad , \quad G_a(z_1) \widetilde{\overline G}{}^b(z_2)  = \mbox{regular}
 \;. \label{linal1}
\eea
Here the central charges $c$ and $c_1$ are the same as in \p{opecoeff} and
the currents $G_a(z),\widetilde{\overline G}{}^a(z)$ and ${\bar Q}^a(z)$ are
bosonic (fermionic) for $\epsilon = 1(-1)$.

In order to prove that the linear algebra  $(Q)SCA_N^{lin}$ \p{linal1}
contains  $U(N)$ (Q)SCA  \p{ope} as a subalgebra,
let us perform the following {\it invertible} nonlinear transformation
to the new basis $\left\{ T(z),U(z),J_a^b(z),G_a(z),{\bar G}^a(z),
{\bar Q}^a(z)\right\}$,
where the "new" current ${\bar G}^a(z)$ is defined as
\bea
{\bar G}^a & = & \widetilde{\overline G}{}^a + y_1 {\bar Q}^a{}''+
 y_2 (J_b^a{\bar Q}^b{}') + y_3 (U{\bar Q}^a{}')+y_4 ({J_b^a}'{\bar Q}^b)+
       y_5 (U'{\bar Q}^a)+y_6 (T{\bar Q}^a)+\nonumber \\
 & & y_7(J_b^cJ_c^a{\bar Q}^b)+
 y_8(J_b^cJ_c^b{\bar Q}^a)+ y_9 (UJ_b^a{\bar Q}^b)+ y_{10} (UU{\bar Q}^a) +
      y_{11}(J_b^cG_c{\bar Q}^b{\bar Q}^a)+  \nonumber \\
 & & y_{12}(J_b^aG_c{\bar Q}^c{\bar Q}^b)+y_{13}(G_b'{\bar Q}^b{\bar Q}^a)+
    y_{14}(G_b{\bar Q}^b{}'{\bar Q}^a)+
    y_{15}(G_b{\bar Q}^b{\bar Q}^a{}')+ \nonumber \\
 & & y_{16}(G_bG_c{\bar Q}^b{\bar Q}^c{\bar Q}^a) +
    y_{17}(UG_b{\bar Q}^b{\bar Q}^a) \; , \label{tr1}
\eea
and the coefficients $y_1-y_{17}$ are defined as
\bea
y_1 & = & 2K \quad , \quad y_2 = 4 \quad , \quad
  y_3=\frac{2(2+\epsilon N)}{N} \quad , \quad
  y_4=\frac{2(K-\epsilon -N)}{K} \;, \nonumber \\
y_5 & = & \frac{(K-\epsilon-N)(2+\epsilon N)}{NK} \quad , \quad
  y_6=-2\epsilon \quad , \quad y_7=\frac{2}{K} \quad , \quad
  y_8=\frac{2}{\epsilon K} \quad , \quad y_9=\frac{2(2+\epsilon N)}{NK}
        \; , \nonumber \\
y_{10} & = & \frac{(\epsilon+N)(2\epsilon+N)}{N^2K} \quad , \quad
 y_{11} = y_{12}=\frac{2}{K} \quad , \quad
 y_{13}= \frac{2(K-N-2\epsilon )}{K}\quad , \quad y_{14}=4 \; \nonumber \\
y_{15} & = & 2 \quad , \quad y_{16}=\frac{2}{\epsilon K} \quad , \quad
 y_{17}=\frac{2(2+\epsilon N)}{NK} \; . \label{sing}
\eea
Now it is a matter of straightforward (though tedious) calculation to check
that  OPE's for the set of currents
$\left\{ T(z),U(z),J_a^b(z),G_a(z)\right\}$ and ${\bar G}^a(z)$ \p{tr1}
coincide with the basic OPE's of the $U(N)$ (Q)SCA \p{ope}.

Thus, we have shown that the linear algebra $(Q)SCA_N^{lin}$ \p{linal1}
contains  $U(N)$ (Q)SCA as a subalgebra in the nonlinear basis.

We close this Section with a few  comments.

First of all, we would like to stress that the pairs of currents
 $G_a(z)$ and ${\bar Q}^a(z)$ (with conformal weights
equal to $3/2$ and $-1/2$, respectively) in \p{linal1} look like
``ghost--anti-ghost'' fields and so $(Q)SCA_N^{lin}$  algebra \p{linal1}
can be simplified by means of the standard ghost decoupling
transformations
\bea
U & = & {\widetilde U}-\epsilon (G_a{\bar Q}^a) \; , \nonumber \\
J_a^b & = & {\widetilde J}{}_a^b - \epsilon (G_a{\bar Q}^b) +
                \delta_a^b\frac{\epsilon}{N}(G_c{\bar Q}^c) \; , \nonumber \\
T & = & {\widetilde T} +\frac{1}{2}\epsilon (G_a'{\bar Q}^a)
       +\frac{3}{2}\epsilon (G_a{\bar Q}^a{}')
        -\frac{\epsilon (2+ \epsilon N)}{2K} {\widetilde U}{}' \; .
                                              \label{ghosts}
\eea
In the new basis the algebra $(Q)SCA_N^{lin}$ splits into the direct product
of the ghost--anti-ghost  algebra $\Gamma_N=\left\{ {\bar Q}^a, G_b \right\}$
with the OPE's
$$
G_a(z_1) {\bar Q}^b(z_2)  =  \frac{\delta_a^b}{z_{12}} \label{gam}
$$
and the algebra of the  currents $\left\{ {\widetilde T},{\widetilde U},
{\widetilde J}{}_a^b,\widetilde{\overline G}{}^a\right\}$. We
denote the latter as $\widetilde{(Q)SCA}{}_N^{lin}$. It is defined by the
following set of OPE's
\bea
{\widetilde T}(z_1){\widetilde T}(z_2) & = &
 \frac{-6\epsilon K^2 +(N^2+13)K -(N^3-N+6\epsilon )}{2K\; z_{12}^4}+
  \frac{2{\widetilde T}}{z_{12}^2}+
                   \frac{{\widetilde T}'}{z_{12}} \quad , \nonumber \\
{\widetilde U}(z_1){\widetilde U}(z_2) & = &
  \left(\frac{2NK}{2+\epsilon N}\right)\frac{1}{z_{12}^2} \; , \;
{\widetilde T}(z_1){\widetilde J}{}_{a}^{b}(z_2) =
 \frac{{\widetilde J}{}_{a}^{b}}{z_{12}^2}+
                   \frac{{\widetilde J}{}_{a}^{b}{}'}{z_{12}} \; ,
                      \nonumber \\
{\widetilde T}(z_1){\widetilde U}(z_2)  & = &
    \frac{{\widetilde U}}{z_{12}^2}+\frac{{\widetilde U}'}{z_{12}} \;,
               \nonumber \\
{\widetilde T}(z_1)\widetilde{\overline G}{}^a(z_2) & = &
 \left( \frac{3}{2}+\frac{\epsilon (2+\epsilon N)}{2K}\right)
            \frac{\widetilde{\overline G}{}^a}{z_{12}^2}+
                 \frac{\widetilde{\overline G}{}^a}{z_{12}}\; , \nonumber \\
{\widetilde J}{}_a^b(z_1){\widetilde J}{}_c^d(z_2) & = &
  (K-N)\frac{\delta_a^d\delta_c^b-\frac{1}{N}\delta_a^b\delta_c^d}{z_{12}^2}+
 \frac{\delta_c^b {\widetilde J}{}_a^d-
          \delta_a^d {\widetilde J}{}_c^b}{z_{12}} \; , \nonumber \\
{\widetilde U}(z_1)\widetilde{\overline G}{}^a(z_2)  & =  &
     -\frac{\widetilde{\overline G}{}^a}{z_{12}}  \; , \;
{\widetilde J}{}_a^b(z_1)\widetilde{\overline G}{}^c(z_2)  =
       \frac{-\delta_a^c \widetilde{\overline G}{}^b +
 \frac{1}{N}\delta_a^b \widetilde{\overline G}{}^c}{z_{12}} \; , \nonumber \\
\widetilde{\overline G}{}^a(z_1) \widetilde{\overline G}{}^b(z_2)  & = &
  \mbox{regular} \;, \label{linal2}
\eea

\be
(Q)SCA_N^{lin}=\Gamma_N \otimes \widetilde{(Q)SCA}{}_N^{lin} \quad .
\ee

Secondly, note that the linear algebra
$\widetilde{(Q)SCA}{}_N^{lin}$ \p{linal2}
has the same number of currents and the same structure relations as the
maximal linear subalgebra ${\cal H}_N$ \p{glin} of $U(N)$ (Q)SCA \p{ope},
but with the "shifted" central charges and
conformal weights. It is of importance that the central charges and
conformal weights are strictly related as in \p{linal2}.\footnote{Let
us remark that Jacoby identities for the set of currents
$\left\{ {\widetilde T},{\widetilde U},
{\widetilde J}{}_a^b,\widetilde{\overline G}{}^a\right\}$
do not fix neither central charges nor the conformal weight of
$\widetilde{\overline G}{}^a$.}
Otherwise, with  another relation between these parameters, we would
never find the $U(N)$ (Q)SCA \p{ope} in
$(Q)SCA{}_N^{lin}$.
Thus, our starting assumption about the structure of linear algebra for
$U(N)$ (Q)SCA  coming from the classical coset realization approach, proved
to be correct, modulo shifts of central charges and conformal weights.

Thirdly, let us remark that among the $U(N)$ (Q)SCAs there are many
(super)algebras which are well known under other names.
For examples:\footnote{To avoid the singularity
in \p{opecoeff} at $\epsilon=-1,N=2$ one should firstly rescale the current
$U\rightarrow \frac{1}{\sqrt{2+\epsilon N}}U$ and then put $\epsilon=-1,N=2$
\cite{KB}.}
\bea
(Q)SCA ( \epsilon=1,N=1) & \equiv & W_3^{(2)}  \quad \cite{PB}, \nonumber \\
(Q)SCA ( \epsilon=-1,N=1) & \equiv & N=2\; SCA \quad \ \cite{A}, \nonumber \\
(Q)SCA ( \epsilon=-1,N=2) & \equiv & N=4
\; SU(2)\; SCA  \quad \cite{A}. \nonumber
\eea

Finally, let us remind that in the simplest case of $W_3^{(2)}$
algebra \cite{KS}, the linear $\widetilde{QSCA}{}_1^{lin}$
algebra \p{linal2} coincides with the linear algebra $W_3^{lin}$
for $W_3$.
For  general $N$ the situation is more complicated. This will be
discussed in the next Section.

\setcounter{equation}0
\section{Linearizing $W$ algebras.}

The problem of construction of linear algebras for nonlinear ones can be
naturally divided in two steps. As the first step we need to find the
appropriate sets of additional currents which linearize the given
nonlinear algebra. In other words, we must construct the
linear algebra (like $(Q)SCA_N^{lin}$) with the correct relations between
all central charges and conformal weights, which  contains the nonlinear
algebra as a subalgebra in some nonlinear basis. As the second step, we
need to explicitly construct the transformation from the linear basis to a
nonlinear one (like \p{tr1}). While the first step is highly non-trivial,
the second one is purely technical. In principle, we could write down the
most general expression with arbitrary coefficients and appropriate
conformal weights, and then fix all the coefficients from the OPE's of
the nonlinear algebra.

In this Section we will demonstrate that the linear algebra
$QSCA_N^{lin}$ \p{linal1} constructed in the previous Section gives us the
hints how to  find the linear algebras for many other $W$-type algebras
which can be obtained from  the  $GL(N)$ QSCAs
via the secondary Hamiltonian reduction \cite{DFRS}.

\subsection{Secondary linearization.}

The bosonic $GL(N)$ QSCAs (or, in another notation, $W_{N+2}^{(N+1)}$), which
have been linearized in the previous Section,
can be obtained through the Hamiltonian reduction
>from the affine $sl(N+2)$ algebras [7-9]. The constraints on the currents of
$sl(N+2)$ algebra which yield $W_{N+2}^{(N+1)}$  read
\be
\left(
  \begin{array}{cc|cccc}
  U &               T & {\overline G}{}^1 & {\overline G}{}^2 & \ldots &
                                     {\overline G}{}^N \\
  1 &               0 & 0   & 0   & \ldots & 0 \\ \hline
  0 & G_1 &     &     &        &   \\
  0 & G_2 &     &     &        &   \\
  \vdots & \vdots &  \multicolumn{4}{c}{ sl(N) - \frac{\delta_a^b}{N}U}   \\
  0 & G_N &     &     &        &
  \end{array}
\right) \label{ss1}
\ee
The $W_{N+2}^{(N+1)}$ algebras, forming in themselves a particular class
of $W$-algebras with quadratic nonlinearity, are at the same time universal
in the sense that a lot of other
$W$-algebras can be obtained from them via the secondary Hamiltonian
reduction (e.g., $W_N$ algebras, etc.)\cite{DFRS}.

Let us consider a set of possible secondary reductions of $W_{N+2}^{(N+1)}$
algebra \p{ss1}. These are introduced by imposing the constraints
\bea
G_1 = 1 \quad , \quad
      G_2=\ldots =G_N=0 \quad ,& & \label{ss2}\\
 \left. sl(N)\right|_{sl(2)} \quad , & & \label{ss3}
\eea
where we denoted as $\left. sl(N)\right|_{sl(2)}$ the set of constraints
on the $sl(N)$ currents,
associated with an arbitrary embedding of $sl(2)$ algebra into $sl(N)$
subalgebra of $W_{N+2}^{(N+1)}$.

The main conjecture we will keep to in this Section is as follows
\begin{quote}\it
To find the linearizing algebra for a given nonlinear $W$-algebra
related to $W_{N+2}^{(N+1)}$ through the Hamiltonian reduction
\p{ss2},\p{ss3}, one should apply the reduction \p{ss3} to the linear algebra
$\widetilde{QSCA}{}_N^{lin}$ \p{linal2} and then linearize the resulting
algebra. The algebra $\widetilde{QSCA}{}_N^{lin}$ itself is the linearizing
algebra for the reduction \p{ss2}.
\end{quote}

Roughly speaking, we propose to replace the linearization of the algebra
$W$ obtained from the nonlinear algebra $W_{N+2}^{(N+1)}$
through the full set of the
Hamiltonian reduction constraints \p{ss2}-\p{ss3}, by the
linearization of the algebra $\widetilde{W}$ obtained from the
{\it linear} algebra $\widetilde{QSCA}{}_N^{lin}$ by imposing  the relaxed
set \p{ss3}.

At present, we are not aware of the rigorous proof of this
statement, but it works well both in the classical cases
(on the level of Poisson brackets) and in many particular quantum examples.
Of course, the secondary Hamiltonian reduction \p{ss3},
being applied to $\widetilde{QSCA}{}_N^{lin}$, gives rise
to a nonlinear algebra. However, the problem of its linearization can
be reduced to the linearization of reduction \p{ss3} applied to the affine
algebra $sl(N)\subset \widetilde{QSCA}{}_N^{lin}$, which
was constructed in \cite{TB}. The resulting algebra will be just linear
algebra for the nonlinear algebra we started with.

Let us briefly discuss the explicit construction of the
linear algebra $W^{lin}$ which contains the nonlinear
algebra $\widetilde{W}$ obtained from $W_{N+2}^{(N+1)}$
via the Hamiltonian reduction constraints \p{ss2}-\p{ss3}.

Let ${\cal J}$ be a current corresponding to the Cartan element $t_0$ of
$sl(2)$ subalgebra. With respect to the adjoint action of $t_0$
the $sl(N)$ algebra can be decomposed into eigenspaces of  $t_0$
with positive,null and negative  eigenvalues $h_a$
\be
sl(N) = \left( sl(N) \right)_{-} \oplus \left( sl(N) \right)_{0} \oplus
         \left( sl(N) \right)_{+} \equiv
        \begin{array}[t]{c}
            \oplus \\
             h_a \end{array}
            \left( sl(N) \right)_{h_a}
\quad . \label{f2}
\ee
(In this subsection, the latin indices $(a,b)$ run over the whole  $sl(N)$,
Greek indices $(\alpha,\beta )$ run over $\left( sl(N) \right)_{-}$ and
the barred Greek ones $(\bar\alpha,\bar\beta )$ over
$\left( sl(N) \right)_{0} \oplus \left( sl(N) \right)_{+} $ .)
The Hamiltonian reduction associated with the embedding \p{f2} can be
performed by putting the appropriate constraints
\be
J_{\alpha}-\chi_{\alpha}=0 \quad , \quad
  \chi_{\alpha}\equiv \chi (J_{\alpha}) \label{sor1}
\ee
on the currents $J_{\alpha}$ from
$\left( sl(N) \right)_{-}$ [2,7]. These constraints are the first class for
integral gradings\footnote{Let us remind, that the half-integer gradings
can be replaced by integer ones, leading to the same reduction \cite{BS}.},
which means that BRST formalism can be used.

In order to impose the constraints \p{sor1} in the framework of BRST
approach one can introduce the fermionic ghost--anti-ghost pairs
$( b_{\alpha},c^{\alpha} )$  with ghost numbers -1 and 1, respectively,
for each current with the negative eigenvalues $h_{\alpha}$:
\be
  c^{\alpha}(z_1)b_{\beta}(z_2) = \frac{\delta^{\alpha}_{\beta}}{z_{12}}
 \quad ,
\ee
and the BRST charge
\be
Q_{BRST} = \int dz J_{BRST}(z) = \int dz
  \left( (J_{\alpha}-\chi (J_{\alpha}))c^{\alpha} -\frac{1}{2}
   f_{\alpha,\beta }^{\gamma}b_{\gamma}c^{\alpha}c^{\beta}\right) \;,
 \label{brst}
\ee
which coincides with that given in the paper \cite{TB}.
The currents of the algebra $\widetilde{QSCA}_N^{lin}$ and the ghost fields
$b_{\alpha},c^{\alpha}$ form the BRST complex, graded by the ghost number.
The $W$ algebra is defined in this approach as the algebra of operators
generating the null cohomology of the BRST charge of this complex.

Following \cite{TB}, let us introduce the "hatted" currents
${\widehat J}{}_a$ :
\be
{\widehat J}{}_a  =  {\widetilde J}{}_a+
            \sum_{\beta,\gamma}
        f_{a,\beta}^{\gamma}b_{\gamma}c^{\beta} \; ,
                        \label{hat1}
\ee
where $f_{a,\beta}^{\gamma}$
are structure constants of $sl(N)$ in the basis \p{f2}.
As shown in \cite{TB}, the $W$-algebras, associated with
the reductions of the affine $sl(N)$ can be embedded into linear algebras
formed by the currents ${\widehat J}{}_{\overline \alpha}$.
In contrast to the sl(N) algebra, our algebra $\widetilde{QSCA}{}_N^{lin}$
contains, besides the $sl(N)$ currents, three additional ones
${\widetilde T},{\widetilde U},{\widetilde{\overline G}}{}^a$.
Fortunately, the presence of these currents create no new problems
while we construct a linearizing algebra for the reduction of
$\widetilde{QSCA}{}_N^{lin}$ by the BRST charge \p{brst}.
Namely, the improved stress-tensor ${\widehat T}$ with respect to which
$J_{BRST}$ in  eq. \p{brst} is a spin 1 primary current can be easily
constructed
\be
{\widehat T}  =  {\widetilde T} +{\cal J}' +
     \sum_{\alpha} \left\{ -(1+h_{\alpha})
         b_{\alpha}c^{\alpha}{}' - h_{\alpha}b_{\alpha}'c^{\alpha}
                \right\}  \; , \label{TT}
\ee
and so it belongs, together with ${\widetilde U}$, which commutes with
$Q_{BRST}$, to a linear algebra we are searching for.
As regards the current ${\widetilde{\overline G}}{}^i$, one could check
that it extends the complex generated
by the currents ${\widehat J}_a,b_{\alpha},c^{\beta}$ with  preserving
the structure of the BRST subcomplexes of the paper \cite{TB},
and forms, together with non-constrained currents
${\widehat J}_{\overline \alpha}$ and $c^{\alpha}$, a reduced BRST
subcomplex and subalgebra which do not contains the currents with
negative ghost numbers. Hence, like in ref. \cite{TB}, the $W$ algebra
closes not only modulo BRST exact operators, but it also closes in
its own right.
So, it is evident that the currents ${\widehat J}_{\overline \alpha}$
also will be present among the currents of the linearizing algebra in our
case, as well as the currents ${\widetilde{\overline G}}{}^i$.

Thus, the set of currents ${\widehat T},{\widehat J}_{\overline \alpha}$
\p{hat1},\p{TT} and the currents
\be
{\widehat U} \equiv {\widetilde U} \quad ,\quad
 {\widehat{\overline G}} \equiv {\widetilde{\overline G}}{}^i \label{sor2}
\ee
form the linear algebra  $W^{lin}$ for the nonlinear
algebra $W$ obtained from  $W_{N+2}^{(N+1)}$
through the secondary Hamiltonian reduction associated with constraints
\p{ss2}-\p{ss3}.

\subsection{Linearizing $W_N$ algebras.}

In this subsection we apply the general procedure described in the previous
subsection to the case of the principal embedding of $sl(2)$ into $sl(N)$
algebra to construct the linear algebras $W_N^{lin}$ which
contain the nonlinear $W_N$ algebras as subalgebras.

For the principal embedding of $sl(2)$ into $sl(N)$ with the currents
$J_a^b, (1\leq a,b \leq N, Tr(J) = 0)$, the current ${\cal J}$ is defined
to be
\be
  {\cal J}= - \sum_{m=1}^{N-1} m J_{N-m}^{N-m} \quad , \label{cartan}
\ee
and the decomposition of affine algebra $sl(N)$ reads as follows
\bea
\left( sl(N) \right)_{-}  \propto  \left\{ J_a^b , ( 2\leq b \leq N ,
         1\leq a <b ) \right\} & & \nonumber \\
\left( sl(N) \right)_{0} \oplus \left( sl(N) \right)_{+} \propto
         \left\{ J_a^b , ( 1\leq a \leq N-1 ,
          a\leq b \leq N ) \right\}  \quad , \label{deco}
\eea
i.e. $\left( sl(N) \right)_{-}$ consists of those entries of the $N\times N$
current matrix which stand below the main
diagonal, and the remainder  just constitutes the subalgebra
$\left( sl(N) \right)_{0} \oplus \left( sl(N) \right)_{+}$.

Now, using \p{linal2},\p{hat1} -- \p{deco},
we are able
to explicitly write the linear algebra $W_{N+2}^{lin}$ which contains
the $W_{N+2}$ algebra as a subalgebra:
\bea
{\widehat T}(z_1){\widehat T}(z_2) & = &
 \frac{(N+1)\left( 1-(N+2)(N+3)\frac{(K-1)^2}{K}\right)}{2z_{12}^4}+
  \frac{2{\widehat T}}{z_{12}^2}+
                   \frac{{\widehat T}'}{z_{12}} \quad , \nonumber \\
{\widehat U}(z_1){\widehat U}(z_2) & = &
  \left(\frac{2NK}{2+N}\right)\frac{1}{z_{12}^2} \; , \nonumber \\
{\widehat T}(z_1){\widehat J}_a^b(z_2) & = &
       \frac{(N+1-2a)(K-1)\delta_a^b}{z_{12}^3}+
        \frac{(b-a+1){\widehat J}_a^b}{z_{12}^2}+
                   \frac{{\widehat J}_a^b{}'}{z_{12}} \; ,
                      \nonumber \\
{\widehat T}(z_1){\widehat U}(z_2)  & = & -\frac{2N(K-1)}{z_{12}^3}+
    \frac{{\widehat U}}{z_{12}^2}+\frac{{\widehat U}'}{z_{12}} \;,
               \nonumber \\
{\widehat T}(z_1)\widehat{\overline G}{}^i(z_2) & = &
            \frac{(i+2)\widehat{\overline G}{}^i}{z_{12}^2}+
                 \frac{\widehat{\overline G}{}^i{}'}{z_{12}}\; , \nonumber \\
{\widehat J}_a^b(z_1){\widehat J}_c^d(z_2) & = &
  K\frac{\delta_a^d\delta_c^b-
       \frac{1}{N}\delta_a^b\delta_c^d}{z_{12}^2}+
        \frac{\delta_c^b {\widehat J}_a^d-
          \delta_a^d {\widehat J}_c^b}{z_{12}} \; , \nonumber \\
{\widehat U}(z_1)\widehat{\overline G}{}^i(z_2)  & =  &
     -\frac{\widehat{\overline G}{}^i}{z_{12}}  \; , \;
{\widehat J}_a^b(z_1)\widehat{\overline G}{}^i(z_2)  =
       \frac{-\delta_a^i \widehat{\overline G}{}^b +
 \frac{1}{N}\delta_a^b \widehat{\overline G}{}^i}{z_{12}} \; , \nonumber \\
\widehat{\overline G}{}^i(z_1) \widehat{\overline G}{}^j(z_2)  & = &
  \mbox{regular} \;, \label{linal3}
\eea
where the indices run over the following ranges:
$$
{\widehat J}_a^b : ( 1\leq a \leq N-1, a\leq b \leq N) \quad , \quad
\widehat{\overline G}{}^i : (1\leq i \leq N) \quad .
$$
In this non-primary basis the currents $\widehat{\overline G}{}^i $ have the
conformal weights
$3,4,...,N+2$, and the stress-tensor
${\widehat T}$ coincides with the stress-tensor of $W_{N+2}$ algebra.

It is also instructive to rewrite the $W_{N+2}^{lin}$ algebra \p{linal3} in
the primary basis   $\left\{ T,{\widehat U},
{\widehat J}{}_a^b,\widehat{\overline G}{}^i\right\}$, where a new
stress-tensor $T$ is defined as
\be
T= {\widehat T}-\frac{(N+2)(K-1)}{2K} {\widehat U}{}'+
  \frac{K-1}{K}\sum_{m=1}^{N-1}  m \left( {\widehat J}{}_{N-m}^{N-m}
         \right)'
\ee
and the OPE's have the following form
\bea
T(z_1) T(z_2) & = &
 \frac{N+1-6\frac{(K-1)^2}{K}}{2z_{12}^4}+\frac{2T}{z_{12}^2}+
                   \frac{ T'}{z_{12}} \quad , \quad
{\widehat U}(z_1){\widehat U}(z_2)  =
  \left(\frac{2NK}{2+N}\right)\frac{1}{z_{12}^2} \; , \nonumber \\
T(z_1){\widehat J}_a^b(z_2) & = &
        \frac{\left( 1-\frac{a-b}{K}\right) {\widehat J}_a^b}{z_{12}^2}+
                   \frac{{\widehat J}_a^b{}'}{z_{12}} \; ,
                      \nonumber \\
T(z_1){\widehat U}(z_2)  & = &
    \frac{{\widehat U}}{z_{12}^2}+\frac{{\widehat U}'}{z_{12}} \;,
               \nonumber \\
T(z_1)\widehat{\overline G}{}^i(z_2) & = &
            \frac{\left(\frac{3}{2}+\frac{1+2i}{2K}\right)
      \widehat{\overline G}{}^i}{z_{12}^2}+
                \frac{\widehat{\overline G}{}^i{}'}{z_{12}}\; , \nonumber \\
{\widehat J}_a^b(z_1){\widehat J}_c^d(z_2) & = &
  K\frac{\delta_a^d\delta_c^b-
       \frac{1}{N}\delta_a^b\delta_c^d}{z_{12}^2}+
        \frac{\delta_c^b {\widehat J}_a^d-
          \delta_a^d {\widehat J}_c^b}{z_{12}} \; , \nonumber \\
{\widehat U}(z_1)\widehat{\overline G}{}^i(z_2)  & =  &
     -\frac{\widehat{\overline G}{}^i}{z_{12}}  \; , \;
{\widehat J}_a^b(z_1)\widehat{\overline G}{}^i(z_2)  =
       \frac{-\delta_a^i \widehat{\overline G}{}^b +
 \frac{1}{N}\delta_a^b \widehat{\overline G}{}^i}{z_{12}} \; , \nonumber \\
\widehat{\overline G}{}^i(z_1) \widehat{\overline G}{}^j(z_2)  & = &
  \mbox{regular} \;. \label{linal4}
\eea
In this basis  the "chain" structure of the algebras $W_N^{lin}$
becomes most transparent. Namely, if we redefine the currents of
$W_{N+2}^{lin}$  as
\bea
{\cal U}_1 & = & {\widehat U}-N\sum_{m=1}^{N-1}{\widehat J}_m^m \;, \nonumber
\\
{\cal U} & = & \frac{(N+2)(N-1)}{N(N+1)}{\widehat U}+
                 \frac{2}{N+1}\sum_{m=1}^{N-1}{\widehat J}_m^m \;
                                          \nonumber \\
{\cal T} & = & T+\sqrt{\frac{N+2}{12KN^2(N+1)}}{\cal U}_1' \; ,
 \quad \left( \mbox{or} \quad
{\cal T} =  T-\frac{N+2}{2KN^2(N+1)}({\cal U}_1{\cal U}_1)\right) \; ,
\nonumber \\
{\cal J}_a^b & = & {\widehat J}_a^b-
            \frac{\delta_a^b}{N-1}
             \sum_{m=1}^{N-1}{\widehat J}_m^m \;,
         (1\leq a \leq N-2, a\leq b \leq N-1 ) \; ,\nonumber \\
{\cal S}_a & = & {\widehat J}_a^N \;, (1\leq a \leq N-1) \; ,\nonumber \\
{\overline{\cal G}}{}^i & = & {\widehat{\overline G}}{}^i
             \;, (1\leq i \leq N-1)\; , \nonumber \\
{\overline{\cal Q}} & = & {\widehat{\overline G}}{}^N  \; , \label{linal5}
\eea
then the subset
${\cal T},{\cal U},{\cal J}_a^b,{\overline{\cal G}}{}^i $  generates
the algebra $W_{N+1}^{lin}$ in the form \p{linal4}. Thus, the
$W_{N+2}^{lin}$ algebras constructed have the following structure
\be
W_{N+2}^{lin}=\left\{ W_{N+1}^{lin}, {\cal U}_1,{\cal S}_a,
        {\overline{\cal Q}} \right\}
\ee
and therefore there exists the following chain of embeddings
\be
\ldots W_{N}^{lin} \subset W_{N+1}^{lin} \subset  W_{N+2}^{lin} \ldots \quad .
       \label{chain}
\ee
Let us stress that the nonlinear $W_{N+2}$ algebras do not possess the
chain structure like \p{chain}, this property is inherent only to their
linearizing algebras $W_{N+2}^{lin}$.

By this we finished the construction of linear algebras $W_{N+2}^{lin}$
which contain  $W_{N+2}$  as subalgebras in a  nonlinear basis.
Let us repeat once more that the explicit expression for the transformations
>from the currents of $W_{N+2}^{lin}$  algebra to those  forming
$W_{N+2}$ algebra
is a matter of straightforward calculation once we know the exact
structure of the linear algebra.

Finally, let us stress that knowing the structure of the linearized
algebras $W_{N+2}^{lin}$  helps us to reveal some interesting properties of
the $W_{N+2}$ algebras and their representations.

First of all, each realization of $W_{N+2}^{lin}$ algebra gives rise to a
realization of $W_{N+2}$.
Hence, the relation
between linear and nonlinear algebras opens a way to find new non-standard
realizations of $W_{N+2}$ algebras. As was shown in \cite{BO} for the
particular
case of $W_3$, these new realizations \cite{KS} can be useful for solving
the problem of embedding Virasoro string into the $W_3$ one.

Among many interesting realizations of $W_{N+2}^{lin}$ there is one very
simple particular realization which can be described as follows.
A careful inspection of the OPE's \p{linal4} shows that the currents
\be
{\widehat{\overline G}}{}^i \; , \;
{\widehat J}_a^b : ( 1\leq a \leq N-1, a < b \leq N)
\ee
are null fields and so they can be consistently put equal to zero.
In this case the algebra $W_{N+2}^{lin}$ will contain only Virasoro
stress tensor $T$ and $N$ $U(1)$-currents $\left\{ {\widehat U},
{\widehat J}_1^1 , \ldots {\widehat J}_{N-1}^{N-1}\right\}$. Of course,
there exists the basis, where all these currents commute with each other.
The currents of $W_{N+2}$ algebra are realized in this basis
in terms of
arbitrary stress tensor $T_{Vir}$ with the central charge $c_{Vir}$
\be
c_{Vir} =1-6\frac{(K-1)^2}{K}
\ee
and $N$ decoupled commuting $U(1)$ currents.
Surprisingly, the values of $c_{Vir}$ corresponding to the
minimal models of Virasoro algebra \cite{min}  at
\be
K=\frac{p}{q} \Rightarrow c_{Vir}=1-6\frac{(p-q)^2}{pq}
\ee
induce the central charge $c_{W_{N+2}}$ of the minimal models for
$W_{N+2}$ algebra \cite{FL}
\be
c_{W_{N+2}}=(N+1)\left(1-(N+2)(N+3)\frac{(p-q)^2}{pq}\right)
\ee
(let us remind that the stress tensor of $W_{N+2}$ coincides with
the stress tensor $\widehat T$ in the non-primary basis \p{linal3}).
For the $W_3$ algebra this property has been discussed in \cite{KS}.

\subsection{Linearizing $W_4$ algebra.}

In this subsection, as an example of our construction, we would like
to present  the explicit formulas concerning the linearization of $W_4$
algebra.

The structure of the linear algebra $W_{4}^{lin}$ in the primary basis
can be immediately read off from the OPE's \p{linal4} by putting $N=2$.
So, the algebra $W_{4}^{lin}$ contains the currents
$\left\{ T,{\widehat U},{\widehat J}{}_1^1,{\widehat J}{}_1^2,
\widehat{\overline G}{}^1,\widehat{\overline G}{}^2\right\}$, with the
conformal weights $\left\{ 2,1,1,\frac{K+1}{K},\frac{3(K+1)}{2K},
\frac{3K+5}{2K}\right\}$, respectively.

Passing to the currents of $W_4$, goes over two steps.

Firstly, we must write down most general, nonlinear in the currents of
$W_4^{lin}$, {\it invertible} expressions for the currents
${\cal T}_W,{\cal W}, {\cal V}$ with the desired conformal weights
(2,3 and 4). It can
be easily done in the nonprimary basis \p{linal3}, where the stress tensor
$\widehat T$  coincides with the stress tensor of $W_4$ algebra.

Secondly, we should calculate the OPE's between the constructed expressions
and demand them to form a closed set.

This procedure completely fixes all coefficients in the expressions for
the currents of $W_4$ algebra in the primary basis in
terms of currents of $W_4^{lin}$ (up to unessential rescalings).
Let us stress that we do not need to
know the explicit structure of $W_4$ algebra. By performing the
second step, we automatically reconstruct the $W_4$ algebra.

Let us present here the results of our calculations for the $W_4$ algebra.
\bea
{\cal T}_W & = & T +\frac{2(K-1)}{K}{\widehat U}'-\frac{K-1}{K}
       {\widehat J}{}_1^1{}' \; , \nonumber \\
{\cal W} & = & {\widehat{\overline G}}{}^1 +\frac{K-1}{K}(T_1-T_2)'+
    \frac{1}{K}\left((T_1-T_2){\widehat U}\right)-
    \frac{K-1}{K}{\widehat J}{}_1^2{}'-
    \frac{1}{K}({\widehat J}{}_1^2{\widehat U}) \; , \nonumber \\
{\cal V} & = & -{\widehat{\overline G}}{}^2 +
    \frac{K-1}{K}{\widehat{\overline G}}{}^1{}'+\frac{1}{2K}\left(
    ({\widehat J}{}_1^2{\widehat J}{}_1^2)+{\widehat J}{}_1^2{}'\right)+
    \frac{1}{K}\left( ({\widehat U}-2{\widehat J}{}_1^1)
           {\widehat{\overline G}}{}^1\right) -
        \frac{1}{K}\left( (T_1-T_2){\widehat J}{}_1^2\right)+ \nonumber \\
 & &   \frac{1}{2K}\left((T_1-T_2)(T_1-T_2)\right)-\frac{2}{K^2}
     \left( {\widehat J}{}_1^1{\widehat J}{}_1^2\right)'+
     \frac{1}{K^2}\left( (T_1+T_2)(2(K-1){\widehat U}'+
         ({\widehat U}{\widehat U}))\right) + \nonumber \\
 & & \frac{K-1}{K^2}\left( (T_1+T_2)'{\widehat U}\right)+
     \frac{(K-1)^2}{2K^2}(T_1+T_2)''-
     \frac{(K-1)(2K-3)(3K-2)}{3K^3}{\widehat U}'''+ \nonumber \\
 & &     \frac{(3-2K)(3K-2)}{4K^3}({\widehat U}''{\widehat U}) -
    \frac{16(6-13K+6K^2)}{K(300-637K+300K^2)}({\cal T}_W{\cal T}_W)-
                 \nonumber \\
 & & \frac{3(60-121K+60K^2)(-6+13K-6K^2)}{4(300-637K+300K^2)}
           {\cal T}_W'' \; , \label{realiz}
\eea
where the auxiliary currents $T_1$ and $T_2$ are defined as
\bea
T_1 & = & T-\frac{1}{K}({\widehat J}{}_1^1{\widehat J}{}_1^1)-
             \frac{1}{2K}({\widehat U}{\widehat U}) \; , \nonumber \\
T_2 & = & \frac{1}{K}({\widehat J}{}_1^1{\widehat J}{}_1^1)-
          \frac{K-1}{K}{\widehat J}{}_1^1{}' \;.
\eea

For the $W_4^{lin}$ algebra \p{linal4}
the currents ${\widehat{\overline G}}{}^1,{\widehat{\overline G}}{}^2 $ and
${\widehat J}{}_1^2$ are null-fields. So we can consistently put them
equal to zero. In this case the expressions \p{realiz} provide us with the
Miura realization of $W_4$ algebra in terms of two currents with conformal
spins 2 ($T_1, T_2$) and with the same central charges,
and one current with spin 1 (${\widehat U}$) which
commute with each other.

\section{Conclusion.}

In this letter we have constructed the linear (super)conformal algebras with
finite numbers of generating currents which contain in some nonlinear basis
a wide class of $W$-(super)algebras, including
$W_N^{(N-1)}$, $U(N)$-superconformal as well as $W_N$ nonlinear algebras.
For the $W_N$ algebras we do not have  a rigorous proof of our conjecture
about the general structure of the linearizing algebras, but
we have shown that it works both
for classical algebras (on the level of Poisson brackets)
and some simplest examples of quantum algebras (e.g.,
for $W_3,W_4$ ). The explicit construction of the linearizing algebras
$W_{N+2}^{lin}$ for $W_{N+2}$  reveals their many
interesting properties: they have a "chain" structure
(i.e. the linear algebras with a given $N$ are subalgebras of those
with a higher $N$), the central charge of the Virasoro subsector of
these linear algebras in the parametrization corresponding to the Virasoro
minimal models, while putting the null-fields equal to zero,
induces the central charge for the minimal models
of $W_N$, etc. This is the reasons why we believe that our conjecture is true.

It is interesting to note that, as we have explicitly demonstrated in the
case of $W_4$ algebra, we do not need to know beforehand
the structure relations of the
nonlinear algebras, which rapidly become very complicated with  growth
of  spins of the involved currents. Once we have constructed the linearizing
algebra, we could algorithmically
reproduce the structure of the corresponding nonlinear
one. So, one of the main open questions now is how much information
about the properties of a given nonlinear algebra we can extract from
its linearizing algebra. The answer to this question could be important
for applications of linearizing algebras to $W$-strings, integrable
systems with $W$-type symmetry, etc. A detailed discussion of this
issue will be given elsewhere. \vspace{0.5cm}

\section*{Note Added.}

     After this paper was completed, we learned of a paper by J.O. Madsen
and E. Ragoucy \cite{mr}, which has some overlap with our work.
They showed that the wide class of $W$-algebras (including $W_n$ ones) can
be linearized in the framework of the secondary hamiltonian reduction.
However, they did not obtain the explicit expressions for the
linearizing algebras (excepting $W_4$ case). The linearization of the
(quasi)superconformal algebras was not considered, because their method
does not allow fields with negative conformal weights.

\section*{Acknowledgments.}

It is a pleasure for us to thank   S. Bellucci,
L. Bonora,  K. Hornfeck, E. Ivanov, V. Ogievetsky, S. Sciuto, A. Semikhatov,
F. Toppan and D. Volkov for many interesting and clarifying discussions.

One of us (A.S.) is also indebt to G. Zinovjev for his interest in this work
and useful discussions.

We are grateful to E. Ivanov for careful reading of the manuscript.

This investigation has been supported in part by the Russian Foundation of
Fundamental Research, grant 93-02-03821, and the International Science
Foundation, grant M9T000.

\end{document}